\documentclass[11pt]{article}

\usepackage{times}

\usepackage[numbers,sort&compress]{natbib}
\usepackage{graphicx}

\usepackage[dvipsnames]{xcolor}

\usepackage{indentfirst}

\usepackage{multicol}

\usepackage{float}

\usepackage{bm}

\usepackage{amsmath}

\usepackage{xkcdcolors}

\usepackage{caption}
\usepackage{setspace}
\usepackage{titling}
\usepackage[labelfont=bf]{caption}

\usepackage{lipsum}  

\newenvironment{sciabstract}{%
\begin{quote}}
{\end{quote}}

\newcounter{lastnote}

\title{Isotope-Selective Strong Field Ionization of Semi-Heavy Water}

\author{
    \normalsize{
        A.J. Howard,$^{1,2,\ast}$ M. Britton,$^{3}$ Z.L. Streeter,$^{4}$ C. Cheng,$^{2,5}$, R.R. Lucchese$^{6}$,
        } \\
    \normalsize{
        C.W. McCurdy,$^{4,6}$ and P.H. Bucksbaum$^{1,2,3,5,\ast}$
        }\\
    \\
    \small{$^{1}$Department of Applied Physics, Stanford University, Stanford, California 94305, USA}\\
    \small{$^{2}$Stanford PULSE Institute, SLAC National Accelerator Laboratory, Menlo Park, California 94025, USA}\\
    \small{$^{3}$Linac Coherent Light Source, SLAC National Accelerator Laboratory, Menlo Park, California 94025, USA}\\
    \small{$^{4}$Department of Chemistry, University of California, Davis, Davis, California 95616, USA}\\
    \small{$^{5}$Department of Physics, Stanford University, Stanford, California 94305, USA}\\
    \small{$^{6}$Chemical Sciences Division, Lawrence Berkeley National Laboratory, Berkeley, California 94720, USA}\\
    \small{$^\ast$To whom correspondence should be addressed; E-mail: ahow@stanford.edu, phbuck@stanford.edu.}
}
\date{
    \small{
        \vspace{-0.5em}\today
        }
}

\begin{document} 



\baselineskip11pt

\twocolumn[{%

    \maketitle 

    \vspace{-3em}
    \begin{sciabstract}
    Semi-heavy water (HOD) is one of the simplest molecules in which the bonds are labelled by isotope.
    We demonstrate that a pair of intense few-femtosecond infrared laser pulses can be used to selectively tunnel ionize along one of the two bonds.
    The first pulse doubly ionizes HOD, inducing rapid bond stretching and unbending.
    Femtoseconds later, the second pulse arrives and further ionization is selectively enhanced along the OH bond.
    These conclusions arise from 3D time-resolved measurements of H$^+$, D$^+$, and O$^+$ momenta following triple ionization.
    \end{sciabstract}

    \noindent
}]


\begin{figure*}[ht]
    \centering
    \includegraphics[width=17.2cm, trim={0cm 0cm 0cm 0cm},clip=true]{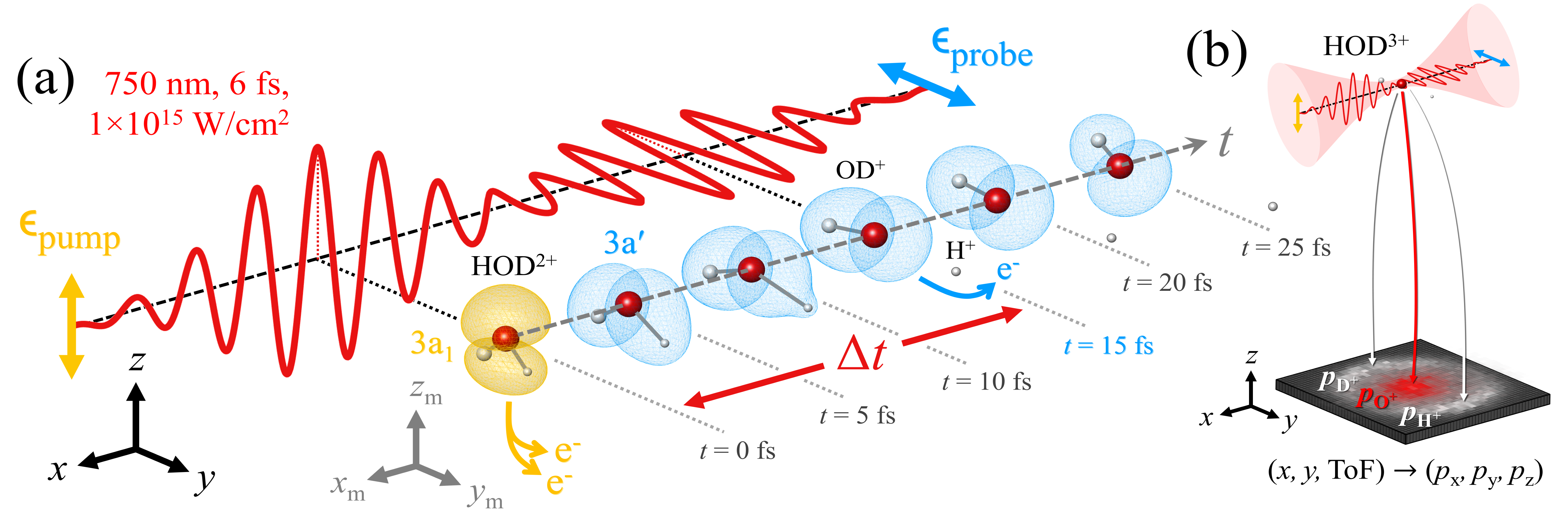}
    \label{fig:Schematic}
    \vspace{-0.1cm}
    \caption{A schematic of the experiment. \textbf{(a)} Two strong-field cross-polarized pulses, denoted pump and probe, sequentially ionize neutral HOD to form HOD$^{2+}$ and HOD$^{3+}$ respectively at an adjustable pump-probe delay ($\Delta t$).
    The pump pulse is polarized along the lab-frame $z$-axis ($\hat{\epsilon}_\mathrm{pump} \parallel \hat{z}$) and the probe pulse is polarized along the lab-frame $y$-axis ($\hat{\epsilon}_\mathrm{probe} \parallel \hat{y}$).
    After the removal of two electrons from the 3a$_1$ molecular orbital of HOD (in yellow), the $z_\mathrm{m}$ axis of the resulting dication is preferentially aligned with the lab-frame $z$-axis.
    The dication then evolves over time ($t$) until the arrival of the probe pulse.
    If the probe pulse arrives at $t\approx$~15~fs, the molecule undergoes Enhanced Ionization (EI) in which an electron is removed from the 3a$^\prime$ (previously 1b$_2$) molecular orbital (in cyan), forming the trication.
    This occurs preferentially along the OH bond.
    \textbf{(b)}~The trication Coulomb explodes into three ionic fragments (H$^+$/D$^+$/O$^+$) which are captured in coincidence by a velocity map imager.
    The $x$ and $y$ positions of each fragment on the detector are used along with their time-of-flight (ToF) in order to reconstruct their 3D momentum distribution ($p_\mathrm{x}$,$p_\mathrm{y}$,$p_\mathrm{z}$) upon formation of the trication.}
\end{figure*}

All chemical processes are mediated by the interatomic motion of electronic charge.
New strategies to view and control the motion of electrons in molecules are improving our understanding of the earliest stages of chemical change \citep{calegari_open_2023,johnston_optics_2023,ivanov_concluding_2021,krausz_attosecond_2009}.
Atomic-site-selective photoionization is one such method which has been indispensable for studies of charge migration and charge transfer \citep{young_roadmap_2018,lepine_attosecond_2014,calegari_charge_2016}.
XUV pulses from high-harmonics and x-ray pulses from free electron lasers have both been used for site-selective ionization \citep{li_attosecond_2020, walter_time-resolved_2022,rebholz_all-xuv_2021,berrah_femtosecond-resolved_2019}.
Here we demonstrate that the strong-field phenomenon of Enhanced Ionization (EI) driven by intense ultrashort infrared (IR) lasers is also atomic-site-selective.
In this case, the selectivity depends on atomic mass, such that isotopic substitution can be used to remove electrons from selected bonds.

Ultrashort IR lasers are often used to image nuclear dynamics in light molecules via laser-induced Coulomb Explosion Imaging (CEI) \citep{cheng_multiparticle_2023,lam_differentiating_2024,endo_capturing_2020,fehre_enantioselective_2019}.
By this method, nuclear geometry is deduced from the ion fragment momenta following non-site-selective multiple ionization \citep{vager_coulomb_1989,stapelfeldt_wave_1995}.
CEI works well in the limit of ultrashort few-femtosecond pulses because they can multiply ionize before the molecular structure changes \citep{howard_filming_2023,howard_strong-field_2021,cheng_strong-field_2021}.
If multiple ionization is performed sequentially and not instantaneously, the atoms in a molecule will start to move before Coulomb explosion.
Site-selective EI may occur in this circumstance, due to distortions of the electrostatic potential created by nuclear motion and seen by certain bound electrons.
EI has been observed in many small molecules to-date, including H$_2$/D$_2$ \citep{zuo_charge-resonance-enhanced_1995,seideman_role_1995,trump_pulse-width_2000,legare_time-resolved_2003,ergler_time-resolved_2005,ben-itzhak_elusive_2008,xu_experimental_2015}, H$_2$O/D$_2$O \citep{legare_laser_2005,liu_charge_2015,mccracken_geometric_2017,howard_filming_2023}, CO$_2$ \citep{brichta_ultrafast_2007,bocharova_charge_2011,song_dissociative_2022}, and C$_2$H$_2$ \citep{gong_strong-field_2014,erattupuzha_enhanced_2017,burger_time-resolved_2018}.
In each case, EI occurs when a bond stretches to a critical length and aligns with the laser polarization.
The phenomenon thus depends on atomic motion, the speed of which changes with atomic mass.
By inducing EI in a hydrogen-containing molecule that has been selectively deuterated, strong-field tunneling effectively becomes isotope-selective.

Isotopic labeling is a common technique in chemistry to tag otherwise indistinguishable atoms within molecules in order to track their evolution through a chemical process \citep{holmes_isotopic_2007,heazlewood_near-threshold_2011,severt_initial-site_2024}.
Semi-heavy water (HOD) is one of the simplest products.
Isotopic substitution does not alter the ground-state electronic structure, but can alter the ionization-driven nuclear dynamics such that strong-field ionization becomes atomic-site-selective.
To demonstrate this effect, we compare the fragmentation patterns of strong-field ionized H$_2$O, HOD, and D$_2$O.

\begin{figure*}[th]
    \centering
    \includegraphics[width=17.2cm, trim={0cm 0cm 0cm 0cm},clip=true]{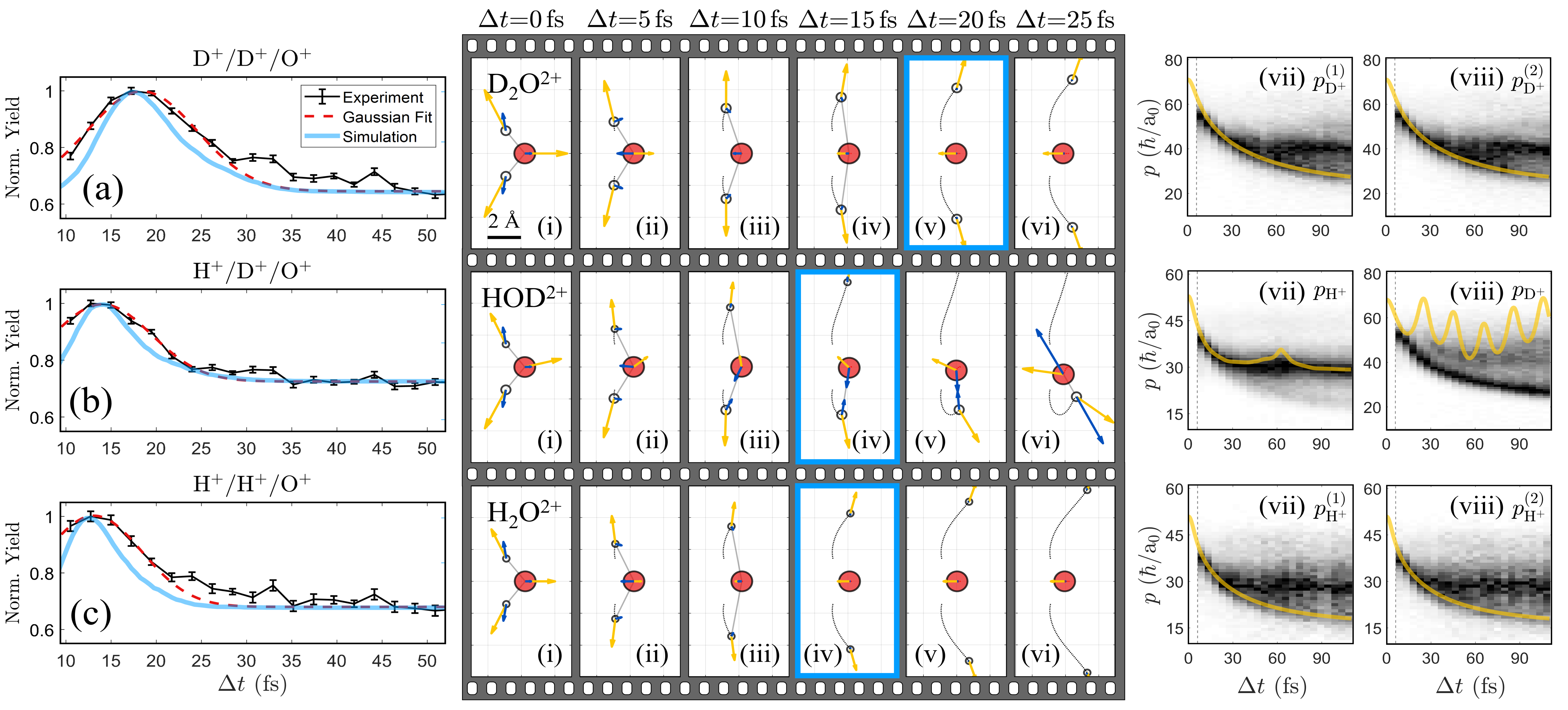}
    \label{fig:Enhancements}
    \vspace{-0.1cm}
    \caption{\textbf{(a-c)} In a solid black line: the yield of the triply-charged three-fold coincidence channel (D$^+$/D$^+$/O$^+$, H$^+$/D$^+$/O$^+$, or H$^+$/H$^+$/O$^+$) as a function of pump-probe delay $\Delta t$ (normalized \textit{in-situ} by the yield of background N$^+$ ions).
    Error bars represent $\sqrt{N}$ noise, where $N$ is the number of coincidences detected.
    In a dotted red line: a Gaussian fit to the early-time experimental data ($\Delta t < 30$~fs for D$^+$/D$^+$/O$^+$, $\Delta t < 21$~fs for H$^+$/D$^+$/O$^+$ and H$^+$/H$^+$/O$^+$).
    In a solid cyan line, a histogram of simulated trajectories that fulfill the EI criteria described in the main text.
    \textbf{(a-c)(i-vi)} A series of film-strip plots displaying the instantaneous molecular geometry at various values of $\Delta t$ for a representative trajectory on the ground-state of the dication (D$_2$O$^{2+}$, HOD$^{2+}$, or H$_2$O$^{2+}$).
    The dark blue vectors represent the instantaneous force on each fragment.
    The yellow vectors represent the asymptotic momentum acquired by each fragment following Coulomb explosion.
    The cyan borders around select frames represent the approximate times at which the ``critical'' geometry for EI is reached.
    \textbf{(a-c)(vii-viii)} The measured momenta of the hydrogen isotope fragments as a function of pump-probe delay.
    In a yellow line: the fragment momenta for the representative trajectory shown in panels (i-vi).}
\end{figure*}

A pair of cross-polarized 6-fs 750-nm pulses with equal peak intensity (I$_0$~=~1$\times$10$^{15}$~W/cm$^2$) sequentially ionize each target molecule.
Two electrons are removed by the ``pump'' pulse.
The dication evolves for a time $\Delta t$.
A ``probe'' pulse then  
removes a third electron.
The trication rapidly Coulomb explodes into three singly-charged fragments, each captured in coincidence by a velocity map imager (VMI) which recovers the full three-dimensional (3D) momentum of each ion.
See Fig.~1 and Supplemental Materials Sections 1.1 $\&$ 1.2 for further experimental details.

The 3D momentum of each fragment versus pump-probe delay directly reflects the sequential dication and trication dynamics. 
This data can be compared to a detailed \textit{ab initio} model to reconstruct an unambiguous state-specific ``movie'' of the real-space interatomic motion in the dication \citep{howard_filming_2023,streeter_dissociation_2018,gervais_h2o2_2009}.
This model assumes the pump pulse promotes the Wigner phase-space distribution of the ground vibrational state of the neutral molecule to one of nine dication potential energy surfaces (PESs) with two vacancies in the valence orbitals.
This distribution is propagated using 4000 classical trajectories on each of the nine PESs for a time, $\Delta t$, before the arrival of the probe pulse.
Upon trication formation, the trajectories are propagated under simple Coulomb repulsion of three singly-charged fragments, yielding asymptotic fragment momenta for each delay.
This theoretical treatment is described in further detail in Supplemental Materials Section 2.1.
 
Previous work has shown that removal of an electron from the 3a$_1$ valence orbital in water leads to rapid unbending and stretching. 
D$_2$O$^{2+}$ with an a$_1$ vacancy evolves to a favorable geometry for EI in about 20 fs \citep{howard_filming_2023}. 
We find that HOD$^{2+}$ and H$_2$O$^{2+}$ also display similar enhancements, but at different time delays, and at an appropriate time delay, HOD$^{2+}$ shows evidence for site-selective ionization. 

The three-body coincidence yield versus pump-probe delay for each isotopologue is displayed in Figs.~2(a-c).
A single maximum occurs in each isotopologue with a Gaussian full-width-at-half-maximum (FWHM) of about 8-fs (D$_2$O:~8.6~fs, HOD:~7.7~fs, H$_2$O:~7.5~fs).
The peak enhancement occurs at a delay of 18.4~fs for D$_2$O, 14.0~fs for HOD, and 13.1~fs for H$_2$O. 

The delayed enhancement for D$_2$O (compared to H$_2$O) is consistent with the slower rovibrational dynamics of the OD bond, which should scale as the square root of the reduced mass: $\mu$~=~$(m_1 m_2)/(m_1 + m_2)$.
Here, $\sqrt{\mu_\mathrm{OD}/\mu_\mathrm{OH}} \approx 1.4$.
The enhancement in HOD is more complicated: there are \textit{not} two peaks corresponding to the different timescales of the OH and OD bonds. 
Rather, the enhancement delay is similar to H$_2$O. 
These observations point to isotope-selective EI. 

The predicted enhancement delay is also plotted in Figs.~2(a-c).
This was estimated using a histogram of our simulated trajectories that fulfill the following enhancement criteria: 
(1) the trajectory is on a PES for a dicationic state with at least one vacancy in the 3a$_1$ orbital, and 
(2) the molecular geometry is in the range of~1.5~$\text{\AA}<$~r$_\mathrm{12}~$\&$~$~r$_\mathrm{23}<$~3.5~$\text{\AA}$, and $\theta_\mathrm{123}>$160$^\circ$, where r$_\mathrm{12}$ and r$_\mathrm{23}$ represent the two bond lengths and $\theta_\mathrm{123}$ represents the bend angle.
Good agreement with our measurement suggests that the mechanism for EI, regardless of isotopologue, results from similar orbital vacancies and ``critical'' geometries.

A representative trajectory is displayed for each isotopologue in Figs.~2(a-c)(i-vi).
The dynamics in D$_2$O$^{2+}$ and H$_2$O$^{2+}$ differ only in speed, both displaying unbending and symmetric stretching that produces a linearized molecule with bond lengths of $>$~2~$\text{\AA}$ during the peak enhancement.
In HOD$^{2+}$, the hydrogen dissociates promptly but the deuterium remains bound.
As seen in Figs.~2(b)(i-vi), after about 10~fs, the attractive force between the oxygen and hydrogen is very weak compared to the force between the oxygen and deuterium.
Consequently, the OD bond never stretches past 2~$\text{\AA}$, but oscillates between approximately 0.9 and 1.5~$\text{\AA}$ with a period of about 20~fs.

The prominence of asymmetric two-body dissociation is seen in Fig.~2(b)(vii), where the momentum of the hydrogen detected in the H$^+$/D$^+$/O$^+$ coincidence predominantly asymptotes to about 30~$\hbar$/a$_0$ at late delays.
This is in contrast to the lower momentum asymptote at about 20~$\hbar$/a$_0$ that results from concerted three-body dissociation.
The propensity of HOD$^{2+}$ to dissociate preferentially into H$^+$/OD$^+$ has been the subject of many prior studies, both theoretical \citep{gervais_h2o2_2009,dey_effect_2016} and experimental \citep{richardson_spectrum_1986,legendre_isotopic_2005,sayler_preference_2006,mathur_selective_2015,jahnke_inner-shell-ionization-induced_2021,guillemin_isotope_2023}.

The larger stretch of the OH bond greatly affects the measured enhancement.
Figs.~3(a,b) show the momentum distribution for triply-charged three-body decay of D$_2$O and HOD at the peak of their respective enhancements.
Note the up-down asymmetry of the distribution for HOD$^{3+}$ as compared to D$_2$O$^{3+}$.
Two separable effects contribute to this asymmetry: firstly, when experiencing the same forces, the asymptotic momentum of D$^+$ will be $\sim$1.4$\times$ larger than H$^+$ due to the mass disparity. 
Secondly, the instantaneous geometry of the molecule is asymmetrically stretched, such that the Coulomb repulsion felt by the D$^+$ is greater than the repulsion felt by the H$^+$. 

Using our simulated trajectories, we can separate the two effects described above.
To do so, we calculate which trajectories best reproduce the experimental momenta of each fragment during the peak of the enhancement.
These filtered trajectories are superimposed in cyan in Figs.~3(a,b).
The geometries that correspond to this subset of trajectories are plotted in Figs.~3(c,d), again in the molecular frame. 
The weighted mean of these geometries gives us the approximate molecular geometry during the enhancement, plotted as a black circle for each atom in Figs.~3(c,d).
In~D$_2$O: $\langle r_\mathrm{OD} \rangle$ = 2.5~$\text{\AA}$, $\langle \theta_\mathrm{DOD} \rangle$~=~173$^\circ$.
In~HOD: $\langle r_\mathrm{OH} \rangle$~=~2.8~$\text{\AA}$, $\langle r_\mathrm{OD} \rangle$~=~1.9~$\text{\AA}$, $\langle \theta_\mathrm{HOD} \rangle$~=~169$^\circ$.
In~H$_2$O: $\langle r_\mathrm{OH} \rangle$~=~2.4~$\text{\AA}$, $\langle \theta_\mathrm{DOD} \rangle$~=~172$^\circ$.
Further detail on the three-body dissociation channels can be found in Supplemental Materials Section 3.

We now consider how the asymmetry in HOD affects the polarization dependence of the enhancement, and compare this case to D$_2$O.
For both isotopologues, we calculate the in-plane angle between the $z_\mathrm{m}$-axis and the probe polarization:
$\phi$~=~arctan$[(\hat{\epsilon}_\mathrm{probe}\cdot\hat{y}_{\mathrm{m}})/$
$(\hat{\epsilon}_\mathrm{probe}\cdot\hat{z}_{\mathrm{m}})]$.
We then plot the normalized yield versus $\phi$ as a polar plot in Figs.~3(a-d).
Fig.~3(d) immediately reveals how the asymmetry in HOD$^{2+}$ affects the enhancement: the distribution peaks when $\hat{\epsilon}_\mathrm{probe}$ aligns with the OH bond, and not the OD bond.
This observation suggests a rare case of bond-selective strong field ionization.
Notably, alignment along the OH bond coincides with alignment along the D$^+$ momentum, as seen in Fig.~3(b), emphasizing the importance of a sophisticated understanding of the nuclear dynamics, without which one might conclude alignment along the OD bond is preferential.
During the enhancement, the OH bond is stretched to approximately 2.8~$\text{\AA}$ whereas the OD is only stretched to 1.9~$\text{\AA}$.
We posit that this preferential alignment results primarily from a lower tunneling barrier along the OH, as compared to the OD, due to the increased bond distance.

\begin{figure}[htbp]
    \centering
    \includegraphics[width=8.6cm, trim={0cm 0cm 0cm 0cm},clip=true]{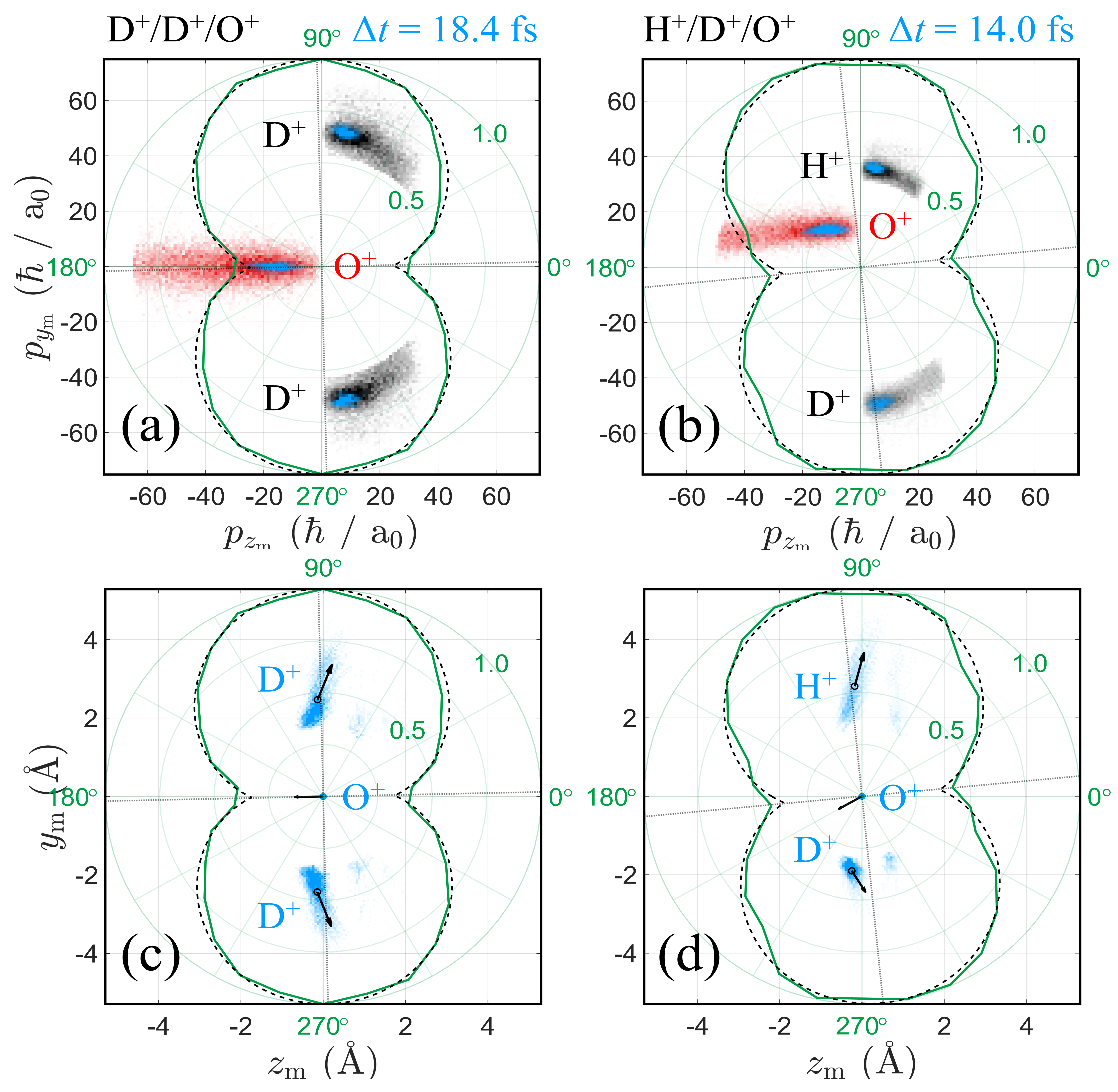}
    \label{fig:Polarization}
    \caption{\textbf{(a,b)} The retrieved molecular-frame momentum distributions ($p_{z_\mathrm{m}},p_{y_\mathrm{m}}$) for each ion in the triply-charged three-body dissociation channel (D$^+$/D$^+$/O$^+$ or H$^+$/D$^+$/O$^+$) in a 4-fs time-window around the maximum enhancement.
    The hydrogen isotopes are plotted in a black color scale and the oxygen is plotted in a red color scale.
    Plotted in a cyan color scale are the simulated trajectories that best reproduce the measured momenta within the specified time-window.
    \textbf{(c,d)} The molecular-frame position-space distributions ($z_\mathrm{m},y_\mathrm{m}$) for the simulated trajectories displayed in (a,b), plotted in a cyan color scale. 
    Here, all positions are relative to the position of the oxygen atom.
    For each ion, the weighted average of the position-space distribution is plotted as a black circle and the weighted average of the velocity distribution is displayed as a black arrow.
    In (a-d), a polar plot is superimposed (in green) representing the normalized yield as a function of the in-plane alignment angle ($\phi$) between the probe pulse polarization axis ($\hat{\epsilon}_\mathrm{probe}$) and the $z_{\mathrm{m}}$ axis.
    A cosine fit of $\phi$ is displayed as a dashed black line, with maxima and minima indicated by dotted black lines.}
\end{figure}

To test our hypothesis, we formed a static tunneling picture.
In this picture, at the moment of EI, the instantaneous geometry of the molecule creates a molecular electrostatic potential (MEP) that the electron to-be-liberated sits in.
This potential is deformed linearly by the presence of a strong laser field directed along the molecular axis, forming a tunneling barrier to the continuum.
If there is sufficient electron density at the barrier and the barrier is sufficiently small, the electron will tunnel out \citep{zuo_charge-resonance-enhanced_1995,seideman_role_1995}.
We simulated the MEP of D$_2$O$^{3+}$ and HOD$^{3+}$ using a multi-configurational self-consistent field (MCSCF) calculation in GAMESS \citep{barca_recent_2020} with a cc-pVTZ Dunning correlation consistent basis set \citep{dunning_gaussian_1989}.
We used the retrieved molecular geometries reported above, and subjected the molecule to an electric field strength equal to approximately half of our peak field strength $\varepsilon$~=~0.08~$E_\mathrm{h}$/$ea_0$~$\sim$~$\varepsilon_0/2$.
From this calculation, we extracted the two-dimensional (2D) MEP of the trication, the 2D electron density of the ionizing orbital in the dication (the 3$a^\prime$), and the ionization potential (IP) of the dication.

The static tunneling picture allows for an intuitive visualization of how EI manifests due to a critical geometry.
Figs.~4(a,b) display the results for D$_2$O and H$_2$O, respectively.
In both cases, the presence of the downhill charge (either D$^+$ or H$^+$) lowers both the internal tunneling barrier (formed between the oxygen and the downhill atomic site) and the external tunneling barrier (formed between the downhill atomic site and the continuum) below the IP.
The lowering of both barriers means that electron density localized at the oxygen can be driven directly out into the continuum. 

In order to explain the asymmetrical polarization dependence of EI for HOD, it is instructive to compare the two above cases (D$_2$O and H$_2$O) with the case of tunneling along the OD atomic site in HOD, shown in Fig.~4(c).
Here, the downhill D lowers the internal tunneling barrier below the IP but not the external barrier at $y_\mathrm{m}$~$\approx$~\mbox{-3~$\text{\AA}$}, which is about \mbox{2.5-$\text{\AA}$} thick.
Tunneling along OH is shown in Fig.~4(d).
Here, the downhill H$^+$ lowers the \textit{external} tunneling barrier below the IP, but not the internal tunneling barrier at $y_\mathrm{m}$~$\approx$~\mbox{1~$\text{\AA}$}, which is only about \mbox{0.8-$\text{\AA}$} thick.
Consequently, electron density is trapped on the uphill site behind a thin barrier--ideal conditions for EI.
In this circumstance, we should expect tunneling ionization to predominantly occur along the OH bond direction, as opposed to the OD, as observed experimentally. 
See Supplemental Materials Section 4 for further analysis of the static tunneling picture.

\begin{figure}[htbp]
    \centering
    \includegraphics[width=8.6cm, trim={0cm 0cm 0cm 0cm}]{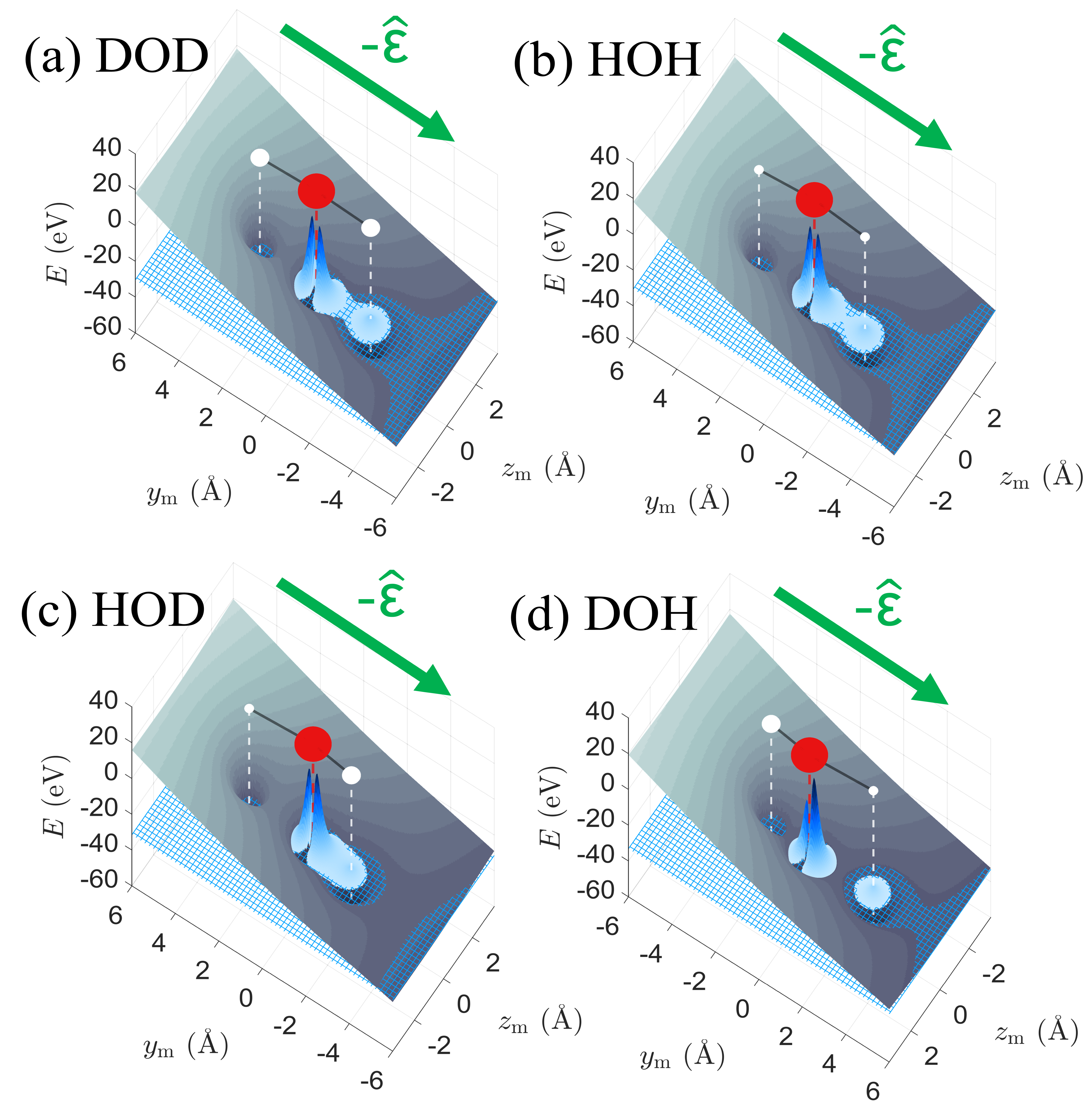}
    \label{fig:Tunneling}
    \vspace{-0.2cm}
    \caption{\textbf{(a)} The molecular electrostatic potential (MEP) of D$_2$O$^{3+}$ for the retrieved geometry that yielded the maximal enhancement, distorted by a static field ($\varepsilon$~=~0.08~$E_\mathrm{h}$/$ea_0$) oriented along the $y_\mathrm{m}$-axis.
    The ionization potential (IP) of the distorted 3a$^\prime$ (previously 1b$_2$) molecular orbital is plotted as a plane with blue grid lines intersecting the MEP at IP~=~\mbox{-30.3~eV}.
    On top of this plane, the electron density of the 3a$^\prime$ orbital is displayed as a surface plot in a cyan color scale. 
    \textbf{(b)} The same as in (a), but for H$_2$O$^{3+}$ (IP~=~\mbox{-30.6~eV}).
    \textbf{(c)} The same as in (a,b), but for HOD$^{3+}$, with the static field oriented along the OD bond (IP~=~\mbox{-31.4~eV}).
    \textbf{(d)} The same as in (a,b), but for HOD$^{3+}$ with the static field oriented along the OH bond (IP~=~-33.1~eV).}
\end{figure}

In summary, through an investigation of Enhanced Ionization in semi-heavy water, we uncovered a unique case in which strong-field ionization becomes isotope-selective.
Using a combination of 3D fragment momentum imaging and ab initio theory, we have constructed a time- and state-resolved ``molecular movie'' of how strong-field ionization ultimately favors tunneling along the OH bond.
Following double ionization, HOD undergoes asymmetric stretching and unbending due to its mass asymmetry.
As a result of this motion, the hydrogen preferentially dissociates as an H$^+$ fragment, whereas the deuterium remains trapped within a bound OD$^+$.
The hydrogen therefore explores much larger internuclear distances than does the deuterium.
The tunneling barrier for electrons in the 3a$^\prime$ molecular orbital is significantly lower at these larger internuclear distances and therefore strong field ionization becomes preferential along the OH bond.
This effect manifests experimentally in the timescale and polarization dependence of EI in HOD$^{2+}$.

Our measurements suggest that by clever use of EI and isotopic substitution, strong-field ionization can be localized along a particular molecular bond, acting as a site-selective probe of charge density.
To verify the efficacy of this technique, further experiments should be performed on other selectively deuterated molecules, especially those larger than HOD.
One such candidate is acetylene (C$_2$H$_2$), which has already been demonstrated to exhibit EI \citep{gong_strong-field_2014,erattupuzha_enhanced_2017,burger_time-resolved_2018}.
Ionization-driven dynamics in the asymmetrically deuterated isotopologue (HC$_2$D) would foreseeably  favor EI along the CH bond.
Extending this experiment to larger molecules may provide a general technique to probe charge density at specific atomic sites within a molecule.
As site selectivity in photoionization is usually the purview of XUV/x-ray sources, this technique may enable the use of table-top IR lasers as a feasible alternative for particular studies of ultrafast charge motion.


\bibliography{ref_bibtex}

\begin{thebibliography}{10}

\bibitem{calegari_open_2023}
F.~Calegari and F.~Martin, ``Open questions in attochemistry,'' {\em Communications Chemistry}, vol.~6, p.~184, Sept. 2023.

\bibitem{johnston_optics_2023}
H.~Johnston, ``Optics pioneers win {Nobel} prize,'' {\em Physics World}, vol.~36, pp.~10--11i, Nov. 2023.

\bibitem{ivanov_concluding_2021}
M.~Ivanov, ``Concluding remarks: {The} age of molecular movies,'' {\em Faraday Discussions}, vol.~228, pp.~622--629, 2021.

\bibitem{krausz_attosecond_2009}
F.~Krausz and M.~Ivanov, ``Attosecond physics,'' {\em Reviews of Modern Physics}, vol.~81, pp.~163--234, Feb. 2009.

\bibitem{young_roadmap_2018}
L.~Young, K.~Ueda, M.~Gühr, P.~H. Bucksbaum, M.~Simon, S.~Mukamel, N.~Rohringer, K.~C. Prince, C.~Masciovecchio, M.~Meyer, A.~Rudenko, D.~Rolles, C.~Bostedt, M.~Fuchs, D.~A. Reis, R.~Santra, H.~Kapteyn, M.~Murnane, H.~Ibrahim, F.~Légaré, M.~Vrakking, M.~Isinger, D.~Kroon, M.~Gisselbrecht, A.~L’Huillier, H.~J. Wörner, and S.~R. Leone, ``Roadmap of ultrafast x-ray atomic and molecular physics,'' {\em Journal of Physics B: Atomic, Molecular and Optical Physics}, vol.~51, p.~032003, Feb. 2018.

\bibitem{lepine_attosecond_2014}
F.~Lépine, M.~Y. Ivanov, and M.~J.~J. Vrakking, ``Attosecond molecular dynamics: fact or fiction?,'' {\em Nature Photonics}, vol.~8, pp.~195--204, Mar. 2014.

\bibitem{calegari_charge_2016}
F.~Calegari, A.~Trabattoni, A.~Palacios, D.~Ayuso, M.~C. Castrovilli, J.~B. Greenwood, P.~Decleva, F.~Martín, and M.~Nisoli, ``Charge migration induced by attosecond pulses in bio-relevant molecules,'' {\em Journal of Physics B: Atomic, Molecular and Optical Physics}, vol.~49, p.~142001, July 2016.

\bibitem{li_attosecond_2020}
J.~Li, J.~Lu, A.~Chew, S.~Han, J.~Li, Y.~Wu, H.~Wang, S.~Ghimire, and Z.~Chang, ``Attosecond science based on high harmonic generation from gases and solids,'' {\em Nature Communications}, vol.~11, p.~2748, June 2020.

\bibitem{walter_time-resolved_2022}
P.~Walter, T.~Osipov, M.-F. Lin, J.~Cryan, T.~Driver, A.~Kamalov, A.~Marinelli, J.~Robinson, M.~H. Seaberg, T.~J.~A. Wolf, J.~Aldrich, N.~Brown, E.~G. Champenois, X.~Cheng, D.~Cocco, A.~Conder, I.~Curiel, A.~Egger, J.~M. Glownia, P.~Heimann, M.~Holmes, T.~Johnson, L.~Lee, X.~Li, S.~Moeller, D.~S. Morton, M.~L. Ng, K.~Ninh, J.~T. O'Neal, R.~Obaid, A.~Pai, W.~Schlotter, J.~Shepard, N.~Shivaram, P.~Stefan, X.~Van, A.~L. Wang, H.~Wang, J.~Yin, S.~Yunus, D.~Fritz, J.~James, and J.-C. Castagna, ``The time-resolved atomic, molecular and optical science instrument at the {Linac} {Coherent} {Light} {Source},'' {\em Journal of Synchrotron Radiation}, vol.~29, pp.~957--968, July 2022.

\bibitem{rebholz_all-xuv_2021}
M.~Rebholz, T.~Ding, V.~Despré, L.~Aufleger, M.~Hartmann, K.~Meyer, V.~Stooß, A.~Magunia, D.~Wachs, P.~Birk, Y.~Mi, G.~D. Borisova, C.~D.~C. Castanheira, P.~Rupprecht, G.~Schmid, K.~Schnorr, C.~D. Schröter, R.~Moshammer, Z.-H. Loh, A.~R. Attar, S.~R. Leone, T.~Gaumnitz, H.~J. Wörner, S.~Roling, M.~Butz, H.~Zacharias, S.~Düsterer, R.~Treusch, G.~Brenner, J.~Vester, A.~I. Kuleff, C.~Ott, and T.~Pfeifer, ``All-{XUV} {Pump}-{Probe} {Transient} {Absorption} {Spectroscopy} of the {Structural} {Molecular} {Dynamics} of {Di}-iodomethane,'' {\em Physical Review X}, vol.~11, p.~031001, July 2021.

\bibitem{berrah_femtosecond-resolved_2019}
N.~Berrah, A.~Sanchez-Gonzalez, Z.~Jurek, R.~Obaid, H.~Xiong, R.~J. Squibb, T.~Osipov, A.~Lutman, L.~Fang, T.~Barillot, J.~D. Bozek, J.~Cryan, T.~J.~A. Wolf, D.~Rolles, R.~Coffee, K.~Schnorr, S.~Augustin, H.~Fukuzawa, K.~Motomura, N.~Niebuhr, L.~J. Frasinski, R.~Feifel, C.~P. Schulz, K.~Toyota, S.-K. Son, K.~Ueda, T.~Pfeifer, J.~P. Marangos, and R.~Santra, ``Femtosecond-resolved observation of the fragmentation of buckminsterfullerene following {X}-ray multiphoton ionization,'' {\em Nature Physics}, vol.~15, pp.~1279--1283, Dec. 2019.

\bibitem{cheng_multiparticle_2023}
C.~Cheng, L.~J. Frasinski, G.~Moğol, F.~Allum, A.~J. Howard, D.~Rolles, P.~H. Bucksbaum, M.~Brouard, R.~Forbes, and T.~Weinacht, ``Multiparticle {Cumulant} {Mapping} for {Coulomb} {Explosion} {Imaging},'' {\em Physical Review Letters}, vol.~130, p.~093001, Mar. 2023.

\bibitem{lam_differentiating_2024}
H.~V.~S. Lam, A.~S. Venkatachalam, S.~Bhattacharyya, K.~Chen, K.~Borne, E.~Wang, R.~Boll, T.~Jahnke, V.~Kumarappan, A.~Rudenko, and D.~Rolles, ``Differentiating {Three}-{Dimensional} {Molecular} {Structures} {Using} {Laser}-{Induced} {Coulomb} {Explosion} {Imaging},'' {\em Physical Review Letters}, vol.~132, p.~123201, Mar. 2024.

\bibitem{endo_capturing_2020}
T.~Endo, S.~P. Neville, V.~Wanie, S.~Beaulieu, C.~Qu, J.~Deschamps, P.~Lassonde, B.~E. Schmidt, H.~Fujise, M.~Fushitani, A.~Hishikawa, P.~L. Houston, J.~M. Bowman, M.~S. Schuurman, F.~Légaré, and H.~Ibrahim, ``Capturing roaming molecular fragments in real time,'' {\em Science}, vol.~370, pp.~1072--1077, Nov. 2020.

\bibitem{fehre_enantioselective_2019}
K.~Fehre, S.~Eckart, M.~Kunitski, M.~Pitzer, S.~Zeller, C.~Janke, D.~Trabert, J.~Rist, M.~Weller, A.~Hartung, L.~P.~H. Schmidt, T.~Jahnke, R.~Berger, R.~Dörner, and M.~S. Schöffler, ``Enantioselective fragmentation of an achiral molecule in a strong laser field,'' {\em Science Advances}, vol.~5, p.~eaau7923, Mar. 2019.

\bibitem{vager_coulomb_1989}
Z.~Vager, R.~Naaman, and E.~P. Kanter, ``Coulomb {Explosion} {Imaging} of {Small} {Molecules},'' {\em Science}, vol.~244, pp.~426--431, Apr. 1989.

\bibitem{stapelfeldt_wave_1995}
H.~Stapelfeldt, E.~Constant, and P.~B. Corkum, ``Wave {Packet} {Structure} and {Dynamics} {Measured} by {Coulomb} {Explosion},'' {\em Physical Review Letters}, vol.~74, pp.~3780--3783, May 1995.

\bibitem{howard_filming_2023}
A.~J. Howard, M.~Britton, Z.~L. Streeter, C.~Cheng, R.~Forbes, J.~L. Reynolds, F.~Allum, G.~A. McCracken, I.~Gabalski, R.~R. Lucchese, C.~W. McCurdy, T.~Weinacht, and P.~H. Bucksbaum, ``Filming enhanced ionization in an ultrafast triatomic slingshot,'' {\em Communications Chemistry}, vol.~6, p.~81, Apr. 2023.

\bibitem{howard_strong-field_2021}
A.~J. Howard, C.~Cheng, R.~Forbes, G.~A. McCracken, W.~H. Mills, V.~Makhija, M.~Spanner, T.~Weinacht, and P.~H. Bucksbaum, ``Strong-field ionization of water: {Nuclear} dynamics revealed by varying the pulse duration,'' {\em Physical Review A}, vol.~103, p.~043120, Apr. 2021.

\bibitem{cheng_strong-field_2021}
C.~Cheng, Z.~L. Streeter, A.~J. Howard, M.~Spanner, R.~R. Lucchese, C.~W. McCurdy, T.~Weinacht, P.~H. Bucksbaum, and R.~Forbes, ``Strong-field ionization of water. {II}. {Electronic} and nuclear dynamics en route to double ionization,'' {\em Physical Review A}, vol.~104, p.~023108, Aug. 2021.

\bibitem{zuo_charge-resonance-enhanced_1995}
T.~Zuo and A.~D. Bandrauk, ``Charge-resonance-enhanced ionization of diatomic molecular ions by intense lasers,'' {\em Physical Review A}, vol.~52, pp.~R2511--R2514, Oct. 1995.

\bibitem{seideman_role_1995}
T.~Seideman, M.~Y. Ivanov, and P.~B. Corkum, ``Role of {Electron} {Localization} in {Intense}-{Field} {Molecular} {Ionization},'' {\em Physical Review Letters}, vol.~75, pp.~2819--2822, Oct. 1995.

\bibitem{trump_pulse-width_2000}
C.~Trump, H.~Rottke, M.~Wittmann, G.~Korn, W.~Sandner, M.~Lein, and V.~Engel, ``Pulse-width and isotope effects in femtosecond-pulse strong-field dissociation of {H} 2 + and {D} 2 +,'' {\em Physical Review A}, vol.~62, p.~063402, Oct. 2000.

\bibitem{legare_time-resolved_2003}
F.~Légaré, I.~V. Litvinyuk, P.~W. Dooley, F.~Quéré, A.~D. Bandrauk, D.~M. Villeneuve, and P.~B. Corkum, ``Time-{Resolved} {Double} {Ionization} with {Few} {Cycle} {Laser} {Pulses},'' {\em Phys. Rev. Lett.}, vol.~91, p.~093002, Aug. 2003.
\newblock Publisher: American Physical Society.

\bibitem{ergler_time-resolved_2005}
T.~Ergler, A.~Rudenko, B.~Feuerstein, K.~Zrost, C.~D. Schröter, R.~Moshammer, and J.~Ullrich, ``Time-{Resolved} {Imaging} and {Manipulation} of {H} 2 {Fragmentation} in {Intense} {Laser} {Fields},'' {\em Physical Review Letters}, vol.~95, p.~093001, Aug. 2005.

\bibitem{ben-itzhak_elusive_2008}
I.~Ben-Itzhak, P.~Q. Wang, A.~M. Sayler, K.~D. Carnes, M.~Leonard, B.~D. Esry, A.~S. Alnaser, B.~Ulrich, X.~M. Tong, I.~V. Litvinyuk, C.~M. Maharjan, P.~Ranitovic, T.~Osipov, S.~Ghimire, Z.~Chang, and C.~L. Cocke, ``Elusive enhanced ionization structure for {H} 2 + in intense ultrashort laser pulses,'' {\em Physical Review A}, vol.~78, p.~063419, Dec. 2008.

\bibitem{xu_experimental_2015}
H.~Xu, F.~He, D.~Kielpinski, R.~Sang, and I.~Litvinyuk, ``Experimental observation of the elusive double-peak structure in {R}-dependent strong-field ionization rate of {H2}+,'' {\em Scientific Reports}, vol.~5, p.~13527, Oct. 2015.

\bibitem{legare_laser_2005}
F.~Légaré, K.~F. Lee, I.~V. Litvinyuk, P.~W. Dooley, S.~S. Wesolowski, P.~R. Bunker, P.~Dombi, F.~Krausz, A.~D. Bandrauk, D.~M. Villeneuve, and P.~B. Corkum, ``Laser {Coulomb}-explosion imaging of small molecules,'' {\em Physical Review A}, vol.~71, p.~013415, Jan. 2005.

\bibitem{liu_charge_2015}
H.~Liu, M.~Li, X.-G. Xie, C.~Wu, Y.-K. Deng, C.-Y. Wu, Q.-H. Gong, and Y.-Q. Liu, ``Charge {Resonance} {Enhanced} {Multiple} {Ionization} of {H} $_{\textrm{2}}$ {O} {Molecules} in {Intense} {Laser} {Fields},'' {\em Chinese Physics Letters}, vol.~32, p.~063301, June 2015.

\bibitem{mccracken_geometric_2017}
G.~A. McCracken, A.~Kaldun, C.~Liekhus-Schmaltz, and P.~H. Bucksbaum, ``Geometric dependence of strong field enhanced ionization in {D2O},'' {\em The Journal of Chemical Physics}, vol.~147, p.~124308, Sept. 2017.

\bibitem{brichta_ultrafast_2007}
J.~P. Brichta, S.~J. Walker, R.~Helsten, and J.~H. Sanderson, ``Ultrafast imaging of multielectronic dissociative ionization of {CO} $_{\textrm{2}}$ in an intense laser field,'' {\em Journal of Physics B: Atomic, Molecular and Optical Physics}, vol.~40, pp.~117--129, Jan. 2007.

\bibitem{bocharova_charge_2011}
I.~Bocharova, R.~Karimi, E.~F. Penka, J.-P. Brichta, P.~Lassonde, X.~Fu, J.-C. Kieffer, A.~D. Bandrauk, I.~Litvinyuk, J.~Sanderson, and F.~Légaré, ``Charge {Resonance} {Enhanced} {Ionization} of {CO} 2 {Probed} by {Laser} {Coulomb} {Explosion} {Imaging},'' {\em Physical Review Letters}, vol.~107, p.~063201, Aug. 2011.

\bibitem{song_dissociative_2022}
P.~Song, Y.~Zhu, Y.~Yang, X.~Wang, C.~Meng, J.~Zhao, J.~Liu, Z.~Lv, D.~Zhang, Z.~Zhao, and J.~Yuan, ``Dissociative multiple ionization of carbon dioxide dimers in intense femtosecond laser fields,'' {\em Physical Review A}, vol.~106, p.~023109, Aug. 2022.

\bibitem{gong_strong-field_2014}
X.~Gong, Q.~Song, Q.~Ji, H.~Pan, J.~Ding, J.~Wu, and H.~Zeng, ``Strong-{Field} {Dissociative} {Double} {Ionization} of {Acetylene},'' {\em Physical Review Letters}, vol.~112, p.~243001, June 2014.

\bibitem{erattupuzha_enhanced_2017}
S.~Erattupuzha, C.~L. Covington, A.~Russakoff, E.~Lötstedt, S.~Larimian, V.~Hanus, S.~Bubin, M.~Koch, S.~Gräfe, A.~Baltuška, X.~Xie, K.~Yamanouchi, K.~Varga, and M.~Kitzler, ``Enhanced ionisation of polyatomic molecules in intense laser pulses is due to energy upshift and field coupling of multiple orbitals,'' {\em Journal of Physics B: Atomic, Molecular and Optical Physics}, vol.~50, p.~125601, June 2017.

\bibitem{burger_time-resolved_2018}
C.~Burger, A.~Atia-Tul-Noor, T.~Schnappinger, H.~Xu, P.~Rosenberger, N.~Haram, S.~Beaulieu, F.~Légaré, A.~S. Alnaser, R.~Moshammer, R.~T. Sang, B.~Bergues, M.~S. Schuurman, R.~De~Vivie-Riedle, I.~V. Litvinyuk, and M.~F. Kling, ``Time-resolved nuclear dynamics in bound and dissociating acetylene,'' {\em Structural Dynamics}, vol.~5, p.~044302, July 2018.

\bibitem{holmes_isotopic_2007}
J.~L. Holmes, K.~J. Jobst, and J.~K. Terlouw, ``Isotopic labelling in mass spectrometry as a tool for studying reaction mechanisms of ion dissociations,'' {\em Journal of Labelled Compounds and Radiopharmaceuticals}, vol.~50, pp.~1115--1123, Oct. 2007.

\bibitem{heazlewood_near-threshold_2011}
B.~R. Heazlewood, A.~T. Maccarone, D.~U. Andrews, D.~L. Osborn, L.~B. Harding, S.~J. Klippenstein, M.~J.~T. Jordan, and S.~H. Kable, ``Near-threshold {H}/{D} exchange in {CD3CHO} photodissociation,'' {\em Nature Chemistry}, vol.~3, pp.~443--448, June 2011.

\bibitem{severt_initial-site_2024}
T.~Severt, E.~Weckwerth, B.~Kaderiya, P.~Feizollah, B.~Jochim, K.~Borne, F.~Ziaee, K.~R. P, K.~D. Carnes, M.~Dantus, D.~Rolles, A.~Rudenko, E.~Wells, and I.~Ben-Itzhak, ``Initial-site characterization of hydrogen migration following strong-field double-ionization of ethanol,'' {\em Nature Communications}, vol.~15, p.~74, Jan. 2024.

\bibitem{streeter_dissociation_2018}
Z.~L. Streeter, F.~L. Yip, R.~R. Lucchese, B.~Gervais, T.~N. Rescigno, and C.~W. McCurdy, ``Dissociation dynamics of the water dication following one-photon double ionization. {I}. {Theory},'' {\em Physical Review A}, vol.~98, p.~053429, Nov. 2018.

\bibitem{gervais_h2o2_2009}
B.~Gervais, E.~Giglio, L.~Adoui, A.~Cassimi, D.~Duflot, and M.~E. Galassi, ``The {H2O2}+ potential energy surfaces dissociating into {H}+/{OH}+: {Theoretical} analysis of the isotopic effect,'' {\em The Journal of Chemical Physics}, vol.~131, p.~024302, July 2009.

\bibitem{dey_effect_2016}
D.~Dey and A.~K. Tiwari, ``Effect of {Vibrational} {Pre}-{Excitation} on the {Dissociation} {Dynamics} of {HOD} $^{\textrm{2+}}$,'' {\em The Journal of Physical Chemistry A}, vol.~120, pp.~2629--2635, May 2016.

\bibitem{richardson_spectrum_1986}
P.~J. Richardson, J.~H.~D. Eland, P.~G. Fournier, and D.~L. Cooper, ``Spectrum and decay of the doubly charged water ion,'' {\em The Journal of Chemical Physics}, vol.~84, pp.~3189--3194, Mar. 1986.

\bibitem{legendre_isotopic_2005}
S.~Legendre, E.~Giglio, M.~Tarisien, A.~Cassimi, B.~Gervais, and L.~Adoui, ``Isotopic effects in water dication fragmentation,'' {\em Journal of Physics B: Atomic, Molecular and Optical Physics}, vol.~38, pp.~L233--L241, July 2005.

\bibitem{sayler_preference_2006}
A.~M. Sayler, M.~Leonard, K.~D. Carnes, R.~Cabrera-Trujillo, B.~D. Esry, and I.~Ben-Itzhak, ``Preference for breaking the {O}–{H} bond over the {O}–{D} bond following {HDO} ionization by fast ions,'' {\em Journal of Physics B: Atomic, Molecular and Optical Physics}, vol.~39, pp.~1701--1710, Apr. 2006.

\bibitem{mathur_selective_2015}
D.~Mathur, K.~Dota, D.~Dey, A.~K. Tiwari, J.~A. Dharmadhikari, A.~K. Dharmadhikari, S.~De, and P.~Vasa, ``Selective breaking of bonds in water with intense, 2-cycle, infrared laser pulses,'' {\em The Journal of Chemical Physics}, vol.~143, p.~244310, Dec. 2015.

\bibitem{jahnke_inner-shell-ionization-induced_2021}
T.~Jahnke, R.~Guillemin, L.~Inhester, S.-K. Son, G.~Kastirke, M.~Ilchen, J.~Rist, D.~Trabert, N.~Melzer, N.~Anders, T.~Mazza, R.~Boll, A.~De~Fanis, V.~Music, T.~Weber, M.~Weller, S.~Eckart, K.~Fehre, S.~Grundmann, A.~Hartung, M.~Hofmann, C.~Janke, M.~Kircher, G.~Nalin, A.~Pier, J.~Siebert, N.~Strenger, I.~Vela-Perez, T.~M. Baumann, P.~Grychtol, J.~Montano, Y.~Ovcharenko, N.~Rennhack, D.~E. Rivas, R.~Wagner, P.~Ziolkowski, P.~Schmidt, T.~Marchenko, O.~Travnikova, L.~Journel, I.~Ismail, E.~Kukk, J.~Niskanen, F.~Trinter, C.~Vozzi, M.~Devetta, S.~Stagira, M.~Gisselbrecht, A.~L. Jäger, X.~Li, Y.~Malakar, M.~Martins, R.~Feifel, L.~P.~H. Schmidt, A.~Czasch, G.~Sansone, D.~Rolles, A.~Rudenko, R.~Moshammer, R.~Dörner, M.~Meyer, T.~Pfeifer, M.~S. Schöffler, R.~Santra, M.~Simon, and M.~N. Piancastelli, ``Inner-{Shell}-{Ionization}-{Induced} {Femtosecond} {Structural} {Dynamics} of {Water} {Molecules} {Imaged} at an {X}-{Ray} {Free}-{Electron} {Laser},'' {\em Physical Review X}, vol.~11, p.~041044, Dec. 2021.

\bibitem{guillemin_isotope_2023}
R.~Guillemin, L.~Inhester, M.~Ilchen, T.~Mazza, R.~Boll, T.~Weber, S.~Eckart, P.~Grychtol, N.~Rennhack, T.~Marchenko, N.~Velasquez, O.~Travnikova, I.~Ismail, J.~Niskanen, E.~Kukk, F.~Trinter, M.~Gisselbrecht, R.~Feifel, G.~Sansone, D.~Rolles, M.~Martins, M.~Meyer, M.~Simon, R.~Santra, T.~Pfeifer, T.~Jahnke, and M.~N. Piancastelli, ``Isotope effects in dynamics of water isotopologues induced by core ionization at an x-ray free-electron laser,'' {\em Structural Dynamics}, vol.~10, p.~054302, Sept. 2023.

\bibitem{barca_recent_2020}
G.~M.~J. Barca, C.~Bertoni, L.~Carrington, D.~Datta, N.~De~Silva, J.~E. Deustua, D.~G. Fedorov, J.~R. Gour, A.~O. Gunina, E.~Guidez, T.~Harville, S.~Irle, J.~Ivanic, K.~Kowalski, S.~S. Leang, H.~Li, W.~Li, J.~J. Lutz, I.~Magoulas, J.~Mato, V.~Mironov, H.~Nakata, B.~Q. Pham, P.~Piecuch, D.~Poole, S.~R. Pruitt, A.~P. Rendell, L.~B. Roskop, K.~Ruedenberg, T.~Sattasathuchana, M.~W. Schmidt, J.~Shen, L.~Slipchenko, M.~Sosonkina, V.~Sundriyal, A.~Tiwari, J.~L. Galvez~Vallejo, B.~Westheimer, M.~Włoch, P.~Xu, F.~Zahariev, and M.~S. Gordon, ``Recent developments in the general atomic and molecular electronic structure system,'' {\em The Journal of Chemical Physics}, vol.~152, p.~154102, Apr. 2020.

\bibitem{dunning_gaussian_1989}
T.~H. Dunning, ``Gaussian basis sets for use in correlated molecular calculations. {I}. {The} atoms boron through neon and hydrogen,'' {\em The Journal of Chemical Physics}, vol.~90, pp.~1007--1023, Jan. 1989.

\end{thebibliography}


\begin{thebibliography}{1}

\bibitem{miranda_simultaneous_2012}
M.~Miranda, T.~Fordell, C.~Arnold, A.~L’Huillier, and H.~Crespo, ``Simultaneous compression and characterization of ultrashort laser pulses using chirped mirrors and glass wedges,'' {\em Optics Express}, vol.~20, p.~688, Jan. 2012.

\bibitem{jagutzki_multiple_2002}
O.~Jagutzki, A.~Cerezo, A.~Czasch, R.~Dorner, M.~Hattas, {Min Huang}, V.~Mergel, U.~Spillmann, K.~Ullmann-Pfleger, T.~Weber, H.~Schmidt-Bocking, and G.~Smith, ``Multiple hit readout of a microchannel plate detector with a three-layer delay-line anode,'' {\em IEEE Transactions on Nuclear Science}, vol.~49, pp.~2477--2483, Oct. 2002.

\bibitem{howard_filming_2023}
A.~J. Howard, M.~Britton, Z.~L. Streeter, C.~Cheng, R.~Forbes, J.~L. Reynolds, F.~Allum, G.~A. McCracken, I.~Gabalski, R.~R. Lucchese, C.~W. McCurdy, T.~Weinacht, and P.~H. Bucksbaum, ``Filming enhanced ionization in an ultrafast triatomic slingshot,'' {\em Communications Chemistry}, vol.~6, p.~81, Apr. 2023.

\bibitem{streeter_dissociation_2018}
Z.~L. Streeter, F.~L. Yip, R.~R. Lucchese, B.~Gervais, T.~N. Rescigno, and C.~W. McCurdy, ``Dissociation dynamics of the water dication following one-photon double ionization. {I}. {Theory},'' {\em Physical Review A}, vol.~98, p.~053429, Nov. 2018.

\bibitem{gervais_h2o2_2009}
B.~Gervais, E.~Giglio, L.~Adoui, A.~Cassimi, D.~Duflot, and M.~E. Galassi, ``The {H2O2}+ potential energy surfaces dissociating into {H}+/{OH}+: {Theoretical} analysis of the isotopic effect,'' {\em The Journal of Chemical Physics}, vol.~131, p.~024302, July 2009.

\bibitem{barca_recent_2020}
G.~M.~J. Barca, C.~Bertoni, L.~Carrington, D.~Datta, N.~De~Silva, J.~E. Deustua, D.~G. Fedorov, J.~R. Gour, A.~O. Gunina, E.~Guidez, T.~Harville, S.~Irle, J.~Ivanic, K.~Kowalski, S.~S. Leang, H.~Li, W.~Li, J.~J. Lutz, I.~Magoulas, J.~Mato, V.~Mironov, H.~Nakata, B.~Q. Pham, P.~Piecuch, D.~Poole, S.~R. Pruitt, A.~P. Rendell, L.~B. Roskop, K.~Ruedenberg, T.~Sattasathuchana, M.~W. Schmidt, J.~Shen, L.~Slipchenko, M.~Sosonkina, V.~Sundriyal, A.~Tiwari, J.~L. Galvez~Vallejo, B.~Westheimer, M.~Włoch, P.~Xu, F.~Zahariev, and M.~S. Gordon, ``Recent developments in the general atomic and molecular electronic structure system,'' {\em The Journal of Chemical Physics}, vol.~152, p.~154102, Apr. 2020.

\end{thebibliography}

\vspace{0.2cm} \noindent \small{\textbf{Acknowledgements:}} A.J.H., M.B., C.C., and P.H.B. were supported by the National Science Foundation.
A.J.H. was additionally supported under a Stanford Graduate Fellowship as the 2019 Albion Walter Hewlett Fellow.
Work at Lawrence Berkeley National Laboratory (LBNL) was performed under the auspices of the U.S. Department of Energy (DOE), Office of Science, Office of Basic Energy Sciences, Chemical Sciences, Geosciences, and Biosciences Division under Contract No. DE-AC02-05CH11231, using the National Energy Research Computing Center (NERSC), a DOE Office of Science User Facility, and the Lawrencium computational cluster resource provided by LBNL.

\end{document}


\maketitle

\vspace{-1.125cm}
\tableofcontents
\clearpage

\vspace{0.5cm}
\addcontentsline{toc}{section}{SUPPLEMENTARY METHODS}
\section*{SUPPLEMENTARY METHODS}
\section{Experimental Methods}
\subsection{Producing Pulse Pairs}
To generate the 6-fs 750-nm pulse pairs, a 40-fs 800-nm 1-kHz Ti:sapphire laser pulse was spectrally broadened in a 1-m long 250-$\mathrm{\mu}$m diameter neon-filled hollow-core fiber (at 45 psi Ne).
At the output of the fiber, the broadened and chirped pulse (now with a central wavelength of 750-nm) was recompressed using a V-shaped chirped mirror block compressor.
The recompressed pulse was split into two pulses of equal intensity and variable interpulse delay using a Mach-Zehnder interferometer.
Each arm contained a polarizer at $\pm$45$^\circ$ to create a cross-polarized pulse pair at the output.
Pulse characterization was performed via dispersion scan \citep{miranda_simultaneous_2012} utilizing two BK7 wedges to apply variable amounts of dispersion.
The pump-probe delay was extracted with high precision ($<$1~fs) from the spectral interference measured between the two beams using a beam sampler and spectrometer.
\subsection{Detecting Coincident Ions}
To sequentially ionize individual water molecules, the beam was sent into a ultra-high vacuum chamber (6~$\times$10$^{-10}$ Torr) and refocused back onto itself using a $f$~=~5~cm in-vacuum spherical metal mirror to form a focal spot of about 7~$\upmu$m.
The chamber was backfilled with a 50/50 mixture of gaseous H$_2$O and D$_2$O to a pressure of approximately ~1.5~$\times$~10$^{-9}$~Torr, such that $<$~1 molecule was in the focus during each laser shot on average.
In this statistical mixture, a large portion of D$_2$O molecules naturally exchange their deuterium atoms with hydrogens, forming HOD.
As depicted schematically in Fig. 1a in the main text, the laser induces sequential multiple ionization in each isotopologue under study, causing the molecule to Coulomb explode into molecular fragments.
These fragments were accelerated toward a detector using a 667~V/cm potential gradient created by series of electrostatic plates held at high voltage.
The detector was comprised of a triple-stack of microchannel plates and a Roentdek delay-line hex-anode \citep{jagutzki_multiple_2002}.
After post-processing of the electrical signals from the detector, this scheme yielded the full 3-dimensional momentum of each ionic fragment captured.
With the laser operating at a repetition rate of 1~kHz, we acquired all ions at an approximate count rate of $\sim$500 counts/s or $\sim$0.5 counts/shot.
\subsection{Reconstructing the Molecular Frame}
Following Coulomb explosion, all fragment momenta are captured in coincidence and their momenta are recorded in lab-frame coordinates ($p_x,p_y,p_z$). 
These coordinates can be rotated into an experimentally recovered molecular frame ($p_{x_\mathrm{m}},p_{y_\mathrm{m}},p_{z_\mathrm{m}}$) by defining a new set of coordinates: $x_\mathrm{m}$, $y_\mathrm{m}$ and $z_\mathrm{m}$. 
Here, $\hat{z}_\mathrm{m}$ is defined as the bisector of the normalized momentum of the two hydrogen isotopes ($\vec{p}_1$ and $\vec{p}_2$), $\hat{x}_\mathrm{m}$ is the normal vector to the plane containing $\vec{p}_1$ and $\vec{p}_2$, and $\hat{y}_\mathrm{m}$ is the in-plane vector perpendicular to $\hat{z}_\mathrm{m}$:

\begin{subequations}
\label{Eq_S1}
\begin{align}
    \hat{z}_\mathrm{m} & = 
    \left(\frac{ \vec{p}_1 }{|\vec{p}_1|} + \frac{ \vec{p}_2 }{|\vec{p}_2|}
    \right) /\hspace{1mm} 2\mathrm{cos}(\beta/2) \label{Eq_1a}\\
    \hat{x}_\mathrm{m} &= \left(\vec{p}_1 \times \vec{p}_2\right) /\hspace{1mm} |\vec{p}_1| \hspace{0.25mm} |\vec{p}_2| \hspace{0.5mm} \mathrm{sin}(\beta) \label{Eq_1b}\\
    \hat{y}_\mathrm{m} &= \hat{z}_\mathrm{m} \times \hat{x}_\mathrm{m} \label{Eq_1c}\\
    \notag\\
    &\hspace{-4mm} \mathrm{where} \hspace{1mm} \beta = \mathrm{arccos}(\vec{p}_1 \cdot \vec{p}_2) \notag
\end{align}
\end{subequations}

\noindent An important limitation of these coordinates is the lack of distinguishability between $+z_\mathrm{m}$ and $-z_\mathrm{m}$. 
Because the bisector of $\hat{p}_1$ and $\hat{p}_2$ is always defined as $+z_\mathrm{m}$, there is nothing distinguishing a molecule that has been inverted about $z_\mathrm{m}$ from one that has not.
This inversion is only apparent after considering the evolution of certain time-resolved observables with pump-probe delay.

\vspace{0.5cm} \section{Theoretical Methods}
\subsection{Modelling Dynamics in HOD$^{2+}$ and H$_2$O$^{2+}$}
\indent The semi-classical treatment of dication dynamics used in this work has been described in detail for D$_2$O$^{2+}$ in a series of previous publications \citep{howard_filming_2023,streeter_dissociation_2018,gervais_h2o2_2009}.
Here, we extended this treatment to model the dynamics of HOD$^{2+}$ and H$_2$O$^{2+}$.
This required altering the following parameters of the model: (1) the atomic masses ($m_1,m_2,m_3$), (2) the normal-mode frequencies ($\omega_1,\omega_2,\omega_3$), and (3) the orthonormal matrix $\mathbf{L}_{ij}$ of the normal-mode eigenvectors, as described in Ref.~\citep{streeter_dissociation_2018}.
The values used for each isotopologue are listed below: \vspace{-1em} \\
\begin{multicols}{2}
\noindent \large{D$_2$O:} \vspace{0.5em} \\
\normalsize
\indent $m_1$ = $m_\mathrm{D}$ = 3670.38 $m_e$\\
\indent $m_2$ = $m_\mathrm{D}$ = 3670.38 $m_e$ \\
\indent $m_3$ = $m_\mathrm{O}$ = 29164.39 $m_e$ \vspace{1em} \\
\indent $\omega_1$ = 1341.82 cm$^{-1}$ (Bend) \vspace{0.5em} \\
\indent \textbf{L}$_{i,j=1}$ = 
$\begin{bmatrix}
0 & -0.28890222 & 0.36282985 \\
0 & 0.28890222 & 0.36282985 \\
0 & 0 & -0.09137130
\end{bmatrix}$ \vspace{0.5em} \\
\indent $\omega_2$ = 2808.06 cm$^{-1}$ (Symmetric Stretch) \vspace{0.5em} \\
\indent \textbf{L}$_{i,j=2}$ = 
$\begin{bmatrix}
0 & -0.40595280 & -0.25821316 \\
0 & 0.40595280 & -0.25821316 \\
0 & 0 & 0.06502571
\end{bmatrix}$ \vspace{0.5em} \\
\indent $\omega_3$ = 2931.57 cm$^{-1}$ (Asymmetric Stretch) \vspace{0.5em} \\
\indent \textbf{L}$_{i,j=3}$ = 
$\begin{bmatrix}
0 & 0.36912871 & 0.27873414 \\
0 & 0.36912871 & -0.27873414 \\
0 & -0.09295782 & 0
\end{bmatrix}$ \vspace{0.5em} \\

\noindent \large{HOD:} \vspace{0.5em} \\
\normalsize
\indent $m_1$ = $m_\mathrm{H}$ = 1837.11 $m_e$\\
\indent $m_2$ = $m_\mathrm{D}$ = 3670.38 $m_e$ \\
\indent $m_3$ = $m_\mathrm{O}$ = 29164.39 $m_e$ \vspace{1em} \\
\indent $\omega_1$ = 1606.19 cm$^{-1}$ (Bend) \vspace{0.5em} \\
\indent \textbf{L}$_{i,j=1}$ = 
$\begin{bmatrix}
0 & -0.45866357 & 0.62427729 \\
0 & 0.22230181 & 0.31246990 \\
0 & 0.00090886 & -0.07868025
\end{bmatrix}$ \vspace{0.5em} \\
\indent $\omega_2$ = 2869.68 cm$^{-1}$ (Symmetric Stretch) \vspace{0.5em} \\
\indent \textbf{L}$_{i,j=2}$ = 
$\begin{bmatrix}
0 & -0.08763001 & 0.01391868 \\
0 & 0.54991665 & -0.37529550 \\
0 & -0.06372129 & 0.04637824
\end{bmatrix}$ \vspace{0.5em} \\
\indent $\omega_3$ = 3951.55 cm$^{-1}$ (Asymmetric Stretch) \vspace{0.5em} \\
\indent \textbf{L}$_{i,j=3}$ = 
$\begin{bmatrix}
0 & 0.79477920 & 0.54956833 \\
0 & 0.00829765 & -0.03164534 \\
0 & -0.05112393 & -0.03064364
\end{bmatrix}$ \vspace{0.5em} \\

\noindent \large{H$_2$O:} \vspace{0.5em} \\
\normalsize
\indent $m_1$ = $m_\mathrm{H}$ = 1837.11 $m_e$\\
\indent $m_2$ = $m_\mathrm{H}$ = 1837.11 $m_e$ \\
\indent $m_3$ = $m_\mathrm{O}$ = 29164.39 $m_e$ \vspace{1em} \\
\indent $\omega_1$ = 1829.53 cm$^{-1}$ (Bend) \vspace{0.5em} \\
\indent \textbf{L}$_{i,j=1}$ = 
$\begin{bmatrix}
0 & -0.38577378 & 0.55535744 \\
0 & 0.38577378 & 0.55535744 \\
0 & 0 & -0.06998605
\end{bmatrix}$ \vspace{0.5em} \\
\indent $\omega_2$ = 3903.28 cm$^{-1}$ (Symmetric Stretch) \vspace{0.5em} \\
\indent \textbf{L}$_{i,j=2}$ = 
$\begin{bmatrix}
0 & -0.58931239 & -0.36354630 \\
0 & 0.58931239 & -0.36354630 \\
0 & 0 & 0.04581404
\end{bmatrix}$ \vspace{0.5em} \\
\indent $\omega_3$ = 3998.52 cm$^{-1}$ (Asymmetric Stretch) \vspace{0.5em} \\
\indent \textbf{L}$_{i,j=3}$ = 
$\begin{bmatrix}
0 & 0.54074491 & 0.40848064 \\
0 & 0.54074491 & -0.40848064 \\
0 & -0.06814481 & 0
\end{bmatrix}$ \\
\end{multicols} \vspace{0.5em}
\indent The normal-mode frequencies and eigenvectors of each isotopologue in its neutral ground state were generated using restricted Hartree-Fock (RHF) theory in GAMESS \citep{barca_recent_2020} with a 6-31G Gaussian basis set.

\newpage
\addcontentsline{toc}{section}{SUPPLEMENTARY RESULTS}
\section*{SUPPLEMENTARY RESULTS}
\section{Triply-charged Three-body Dissociations}
To produce the momentum distributions displayed in Figs.~3(a,b), representing the triply-charged three-body dissociations of each isotopologue (D$^+$/D$^+$/O$^+$, H$^+$/D$^+$/O$^+$, and H$^+$/H$^+$/O$^+$), we utilized both 2-body and 3-body coincidence data.
The 2-body coincidences in which only the hydrogen isotopes were captured (D$^+$/D$^+$, H$^+$/D$^+$, and H$^+$/H$^+$) have the following advantages over the full 3-body coincidences: (1) the momentum resolution for the lighter masses is higher (due to the nature of the detector as a velocity imager), and (2) the detection efficiency is $\sim$100$\%$ higher when detecting two coincident fragments rather than three. 
For this reason, Fig.~3 in the main text uses a filtered subset of the 2-body coincidence data that results from the triply-charged three-body dissociation instead of the full three-body coincidence data. \\
\indent In order to use the two-body coincidence data as a representation of the full three-body coincidence data, it was first necessary to retrieve the momentum of the unseen fragment via momentum conservation, that is: $\vec{p}_3 = -(\vec{p}_1 + \vec{p_2})$, where $\vec{p}_1$ and $\vec{p}_2$ represent the momenta of the detected fragments (D$^+$/D$^+$, H$^+$/D$^+$, or H$^+$/H$^+$), and $\vec{p}_3$ represents the O$^+$ momentum.
We then found the average values for $\vec{p}_1$, $\vec{p}_2$, and $\vec{p}_3$ following triply-charged three-body decay using the full 3-body coincidence data while enforcing a momentum conservation filter ($|\vec{p}_1 + \vec{p}_2 + \vec{p}_3| \leq 20~\hbar/a_0$) in order to discard any instances of false coincidence. 
This filtered 3-body coincidence data, displayed in Fig.~S1, can be used to construct gates in momentum-space that filter out any contributions to the 2-body coincidence resulting from channels other than the intended triply-charged three-body decay.
The momentum gates for each isotopologue (as well as the total number of coincidences in the 2-body and 3-body coincidence channels) are listed below:
\vspace{-1em} \\
\begin{multicols}{2}
\noindent \large{D$^+$/D$^+$/O$^+$:} \vspace{0.5em} \\
\normalsize
\indent $p_1 \geq $~40~$\hbar/a_0$ \\
\indent $p_2 \geq $~40~$\hbar/a_0$ \\
\indent $p_3 \leq $~70~$\hbar/a_0$ \vspace{0.5em}\\
\indent 3-body coincidences: 5,247\\
\indent 2-body coincidences (after filter): 16,144 \\
\\
\noindent \large{H$^+$/D$^+$/O$^+$:} \vspace{0.5em} \\
\normalsize
\indent $p_1 \geq $~31~$\hbar/a_0$ \\
\indent $p_2 \geq $~42~$\hbar/a_0$ \\
\indent $p_3 \leq $~50~$\hbar/a_0$ \vspace{0.5em}\\
\indent 3-body coincidences: 7,126 \\
\indent 2-body coincidences (after filter): 15,926 \\

\noindent \large{H$^+$/H$^+$/O$^+$:} \vspace{0.5em} \\
\normalsize
\indent $p_1 \geq $~30~$\hbar/a_0$ \\
\indent $p_2 \geq $~30~$\hbar/a_0$ \\
\indent $p_3 \leq $~40~$\hbar/a_0$ \vspace{0.5em} \\
\indent 3-body coincidences:  2,331 \\
\indent 2-body coincidences (after filter): 5,534 \vspace{9em} \\
\end{multicols}

\vspace{-1em} \noindent These gates are also reproduced graphically and superimposed on both the full 3-body coincidence data (Fig.~S1) and the filtered 2-body coincidence data (Fig.~S2) as red and black circular arcs.
Note, in comparing Fig.~S2, and Fig.~S1, how the preferential alignment of $\phi$ along the OH bond is consistent between the 3-body and 2-body coincidence data.

\begin{figure}[H]
    \centering
    \includegraphics[width=17.2cm, trim={0cm 0cm 0cm 0cm},clip=true]{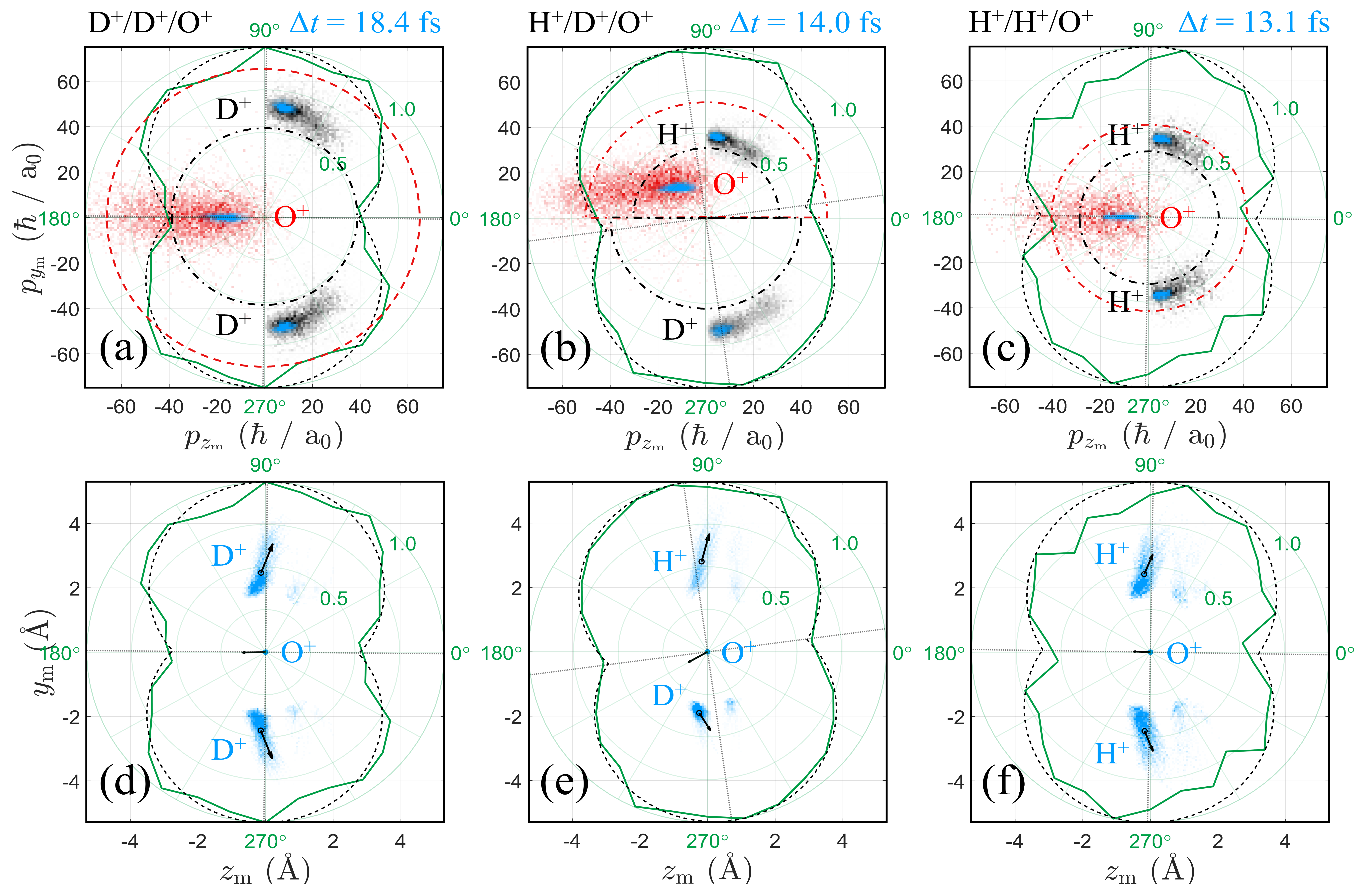}
    \label{fig:S1_Polarization}
    \caption{\textbf{(a-c)} The retrieved molecular-frame momentum distributions ($p_{z_\mathrm{m}},p_{y_\mathrm{m}}$) for each ion in the triply-charged three-fold coincidence channel (D$^+$/D$^+$/O$^+$, H$^+$/D$^+$/O$^+$, or H$^+$/H$^+$/O$^+$) in a 4-fs time-window around the maximum enhancement.
    The hydrogen isotopes are plotted in a black color scale and the oxygen is plotted in a red color scale.
    Plotted in a cyan color scale are the simulated trajectories that best reproduce the measured momenta within the specified time-window.
    In green, a polar plot is superimposed representing the normalized yield as a function of the in-plane alignment angle ($\phi$) between the probe pulse polarization axis ($\hat{\epsilon}_\mathrm{probe}$) and the $z_{\mathrm{m}}$ the molecular axis.
    A cosine fit of $\phi$ is displayed as a dashed black line.
    The maxima and minima of this fit are plotted as dotted black lines.
    The momentum gates used to distinguish the 2-body coincidences that result from this 3-body dissociation channel are displayed as red (for oxygen) and black (for hydrogen) dash-dotted circular arcs.
    \textbf{(d-f)} The molecular-frame position-space distributions ($z_\mathrm{m},y_\mathrm{m}$) for the simulated trajectories displayed in (a-c), plotted in a cyan color scale. 
    Here, all positions are relative to the position of the oxygen atom.
    Superimposed in green is the same polar plot as in (a-c). 
    For each ion, the weighted average of the position-space distribution is plotted as a black circle and the weighted average of the velocity distribution is displayed as a black arrow.}
\end{figure}

\begin{figure}[H]
    \centering
    \includegraphics[width=17.2cm, trim={0cm 0cm 0cm 0cm},clip=true]{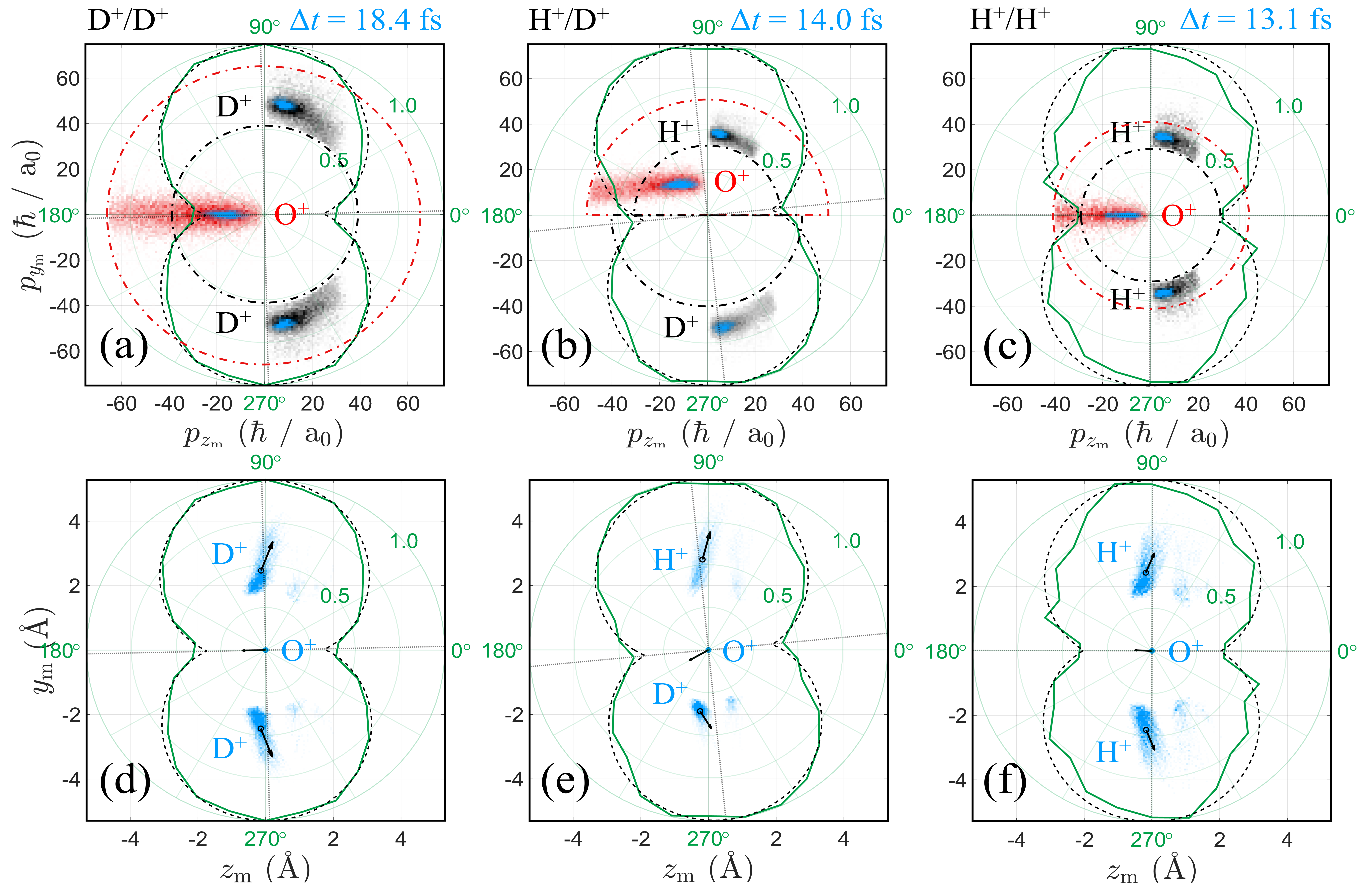}
    \label{fig:S1_Polarization}
    \caption{The same as Fig.~S1 but for the 2-body coincidence channels consisting of simultaneous detection of the hydrogen isotopes (D$^+$/D$^+$, H$^+$/D$^+$, or H$^+$/H$^+$).
    Here, the momentum gates listed in the supplemental text and displayed in Fig.~S1 have been applied to filter out all the data that does not come from triply-charged three-body dissociation.}
\end{figure}

\newpage
\section{Static Tunneling Picture}
\indent The static tunneling picture, displayed in Fig.~4 of the main text may qualitatively explain the difference in the degree of the enhancement for each isotopologue.
Seen in Fig.~1(a-c) of the main text, enhancements of 55$\%$,~37$\%$, and 47$\%$ were found for D$_2$O,~HOD, and H$_2$O respectively.
The disparity between D$_2$O and H$_2$O may be explained by the difference in timescales, that is: D$_2$O spends more time at its critical geometry than H$_2$O.
However, the degree of the enhancement for HOD is still 10$\%$ lower than H$_2$O, and no such disparity in timescales exists. 
The lower degree of enhancement for HOD may arise from the fact that the enhancement can only occur every half-cycle of the field, due to the asymmetric critical geometry.
Additionally, according to Fig.~4(a-d), the threshold for over-the-barrier ionization is higher for HOD than for H$_2$O (or D$_2$O), as evidenced by the presence of an internal barrier for HOD at a field strength of 0.08~$E_\mathrm{h}$/$ea_0$, whereas no such barrier exists for H$_2$O (or D$_2$O). \\

\bibliography{ref_bibtex}